\documentclass[runningheads]{svmult}

\usepackage{makeidx}   % allows index generation
\usepackage{graphicx}  % standard LaTeX graphics tool
\usepackage{subeqnar}  % subnumbers individual equations
\usepackage{multicol}  % used for the two-column index
\usepackage{physprbb}  % modified textarea for proceedings,
\makeindex             % used for the subject index
\begin{document}
\title*{Extragalactic Globular Clusters in the Near-Infrared}
\toctitle{Extragalactic Globular Clusters in 
\protect\newline the Near-Infrared}
% allows explicit linebreak for the table of content
%
%
\titlerunning{Extragalactic Globular Clusters}
% allows abbreviation of title, if the full title is too long
% to fit in the running head
%
\author{Maren Hempel\inst{1}
\and Markus Kissler-Patig\inst{1}
\and Michael Hilker\inst{2}
\and Thomas H. Puzia\inst{3}
\and Jean P.Brodie\inst{4}
\and Paul Goudfrooij\inst{5}
\and Dante Minniti\inst{6}
\and Stephen E.Zepf\inst{7}}
\authorrunning{Maren Hempel et al.}

\institute{European Southern Observatory, Garching 85748, Germany
\and Sternwarte Bonn, Germany
\and Sternwarte der Ludwig-Maximilians Universit\"at, M\"unchen, Germany
\and UCO/Lick observatory, University of California, Santa Cruz, USA 
\and Space Telescope Science Institute, Baltimore, US 
\and P. Universidad Cat\'olica, Santiago, Chile
\and Yale University, New Haven, USA
}
\maketitle              % typesets the title of the contribution

\begin{abstract}
Visual and near-infrared colours are used to identify intermediate-age
globular cluster in early-type galaxies. In NGC~5846 and NGC~4365
colour-colour diagrams (V-I vs. V-K) reveal a sub-population
in the globular cluster systems with an age much younger than the expected old populations
(13-15 Gyr). A comparison with SSP models dates the formation of
this younger population 2-7 Gyr back. Further evidence for
multiple star formation events is given by the age distribution in
both systems and the comparison to models. Our photometric age 
determination was confirmed by spectroscopy for a small sample of the
clusters in NGC~4365. 

\end{abstract}

\section{Introduction}
Globular cluster are known to form during major star formation
episodes \cite{holtz92},\cite{kissler00}. Consequently the star
formation history in galaxies can coarsely be described by the age
structure of the globular cluster system. The detection of
sub-populations and the determination of their ages is therefore of
great interest in galaxy formation studies. But besides the pure
detection of intermediate-age populations the importance of recent
star formation events for the galaxy evolution has to be
considered. Using combined optical and near-infrared photometry
provides the observational tools for this task. Taking advantage of
the different sensitivity to age and metallicity of colours in the
visual and near-infrared bands \cite{puzia02} the degeneracy of both
parameters \cite{worthey94} can be lifted. Photometry as opposite to
spectroscopy allows us to quantify the age distribution by including a
large number of objects within the globular cluster system. The draw
back is photometry being a somewhat less accurate age indicator and
less informative on the chemistry. First results of this approach
towards age determination and identification of multiple star
formation events have been published by Kissler-Patig et
al. \cite{kissler02a} and Puzia et al. \cite{puzia02}. It is highly
encouraging that the photometric results have recently been confirmed
by spectroscopy
\cite{larsen02}. The results presented here are part of this ongoing
project on globular cluster systems of early-type galaxies.

\section{The Project}

%\subsection{Globular Cluster Sample}

Our complete galaxy sample consists of 11 early-type galaxies covering
a wide range of luminosities in various environments, e.g. central
cluster or group galaxies (NGC~1399, M87, NGC~5846) and isolated
ellipticals (NGC~7192, NGC~3115, NGC~1426). In this study we combine
near-infrared data (K$_{s}$-band, obtained with ISAAC) and optical
(B-, V-and I-band, WFPC2/HST archive) photometry. The observed
ISAAC fields were chosen to match the HST archive data where the PC
chip was centred on the galaxy. The data have been reduced in the
standard way using IRAF and SExtractor \cite{bertin96} as described in
\cite{kissler02a} and \cite{puzia02}. 

\section{Results}
\subsection{Colour Distributions}
\begin{figure}[]
\begin{center}
\includegraphics[width=1.0\textwidth]{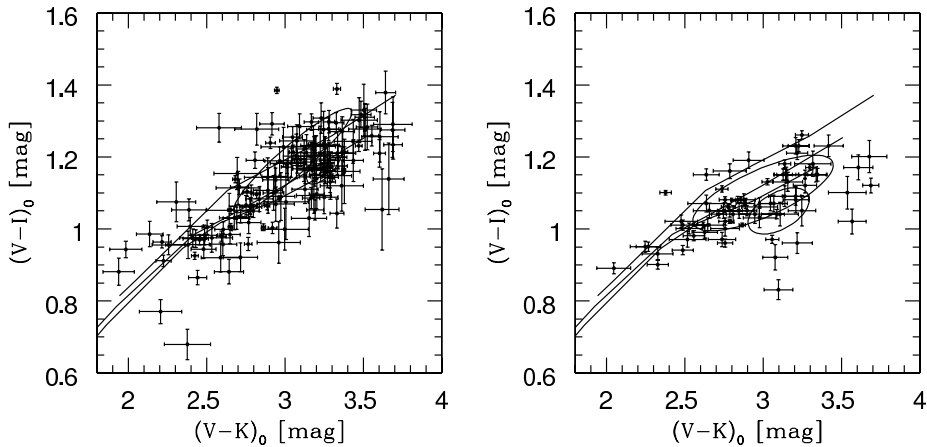}
\end{center}
\caption[]{(V-I) vs. (V-K) colour-colour diagram for NGC~5846 and NGC~4365. The
solid lines mark the isochrones for a 15,~5 and 3 Gyr SSP model. The
ellipses are eye guides to the two identified red
populations.}
\label{vikcolcol}
\end{figure}

Intermediate age populations can be identified by using in
colour-colour diagrams, e.g. (V-I) vs. (V-K). Examples of these are
given in Fig. \ref{vikcolcol}. Hereby it is important that the
different completeness levels for optical and near-infrared colours
are taken into account. A limit in photometric errors (0.15 mag in
V-I) is used as additional selection criterion. In order to exclude
background galaxies we used the photometric FWHM estimated by
SExtractor and compared it with the one we obtained for point
sources.\\ In the galaxies shown here (NGC~4365 and NGC~5846) we find
a distinct second red population (indicated by small ellipses) of
globular clusters which are red in (V-K) with
(V-K)~$\in$~$\langle$2.9,3.2$\rangle$ but intermediate in (V-I) with
(V-I)~$\in$~$\langle$1.0,1.2$\rangle$.

\subsection{Age Distribution}
\begin{figure}[]
\begin{center}
\includegraphics[width=.7\textwidth]{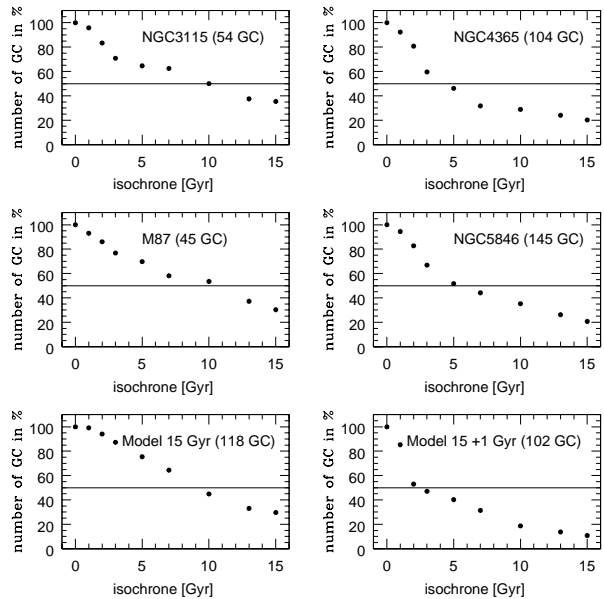}
\end{center}
\caption{Age distribution for GCS in different galaxies (NGC~5846,
NGC~4365, M87 and NGC~3115). Each panel shows the
cumulative age distribution in a given galaxy, the total number of
objects is given in brackets. The two lower panels show the
distributions for artificial data (left: a purely 15 Gyr old
and right: 50$\%$ 15 Gyr and 50$\%$ 1 Gyr). }
\label{galaxies_gc}
\end{figure}

Our age determination of globular clusters relies on the comparison of
SSP isochrones with colour-colour diagrams of globular cluster
systems. Various single stellar population models (SSP) built by
Bruzual \& Charlot \cite{bruzual00}, Maraston \cite{maraston00},
Vazdekis \cite{vazdekis99} or Kurth et al. \cite{kurth99} are
available. For the work discussed here the model isochrones by Bruzual
\& Charlot are used. Given the uncertainty in the isochrone modelling
($\pm$3Gyr) we do not intend to determine absolute ages. The goal is
rather to identify possible sub-populations and to obtain their
relative ages.  A first attempt to distinguish the age structure of
globular cluster systems in different environment is given in
Fig.~\ref{galaxies_gc}. This figure shows our quantitative method to
detect intermediate age populations in globular cluster systems,
supplementing the purely visual inspection of colour-colour
diagrams. Ages for each cluster are determined in the colour-colour
diagrams with respect to the model isochrones. The 50$\%$ level is
marked by a solid line. For comparison we modelled the colour-colour diagrams for
a pure 15 Gyr old population and a mixed
system respectively, consisting of 50$\%$ 15 Gyr and 50$\%$ 1 Gyr old
globular clusters. The results were than evaluated in the same way as
the observed data.

As shown in Fig.\ref{galaxies_gc}, M87 and NGC~3115 have age
distributions compatible to a pure old systems \cite{puzia02},
\cite{jordan02} (see lower left panel) with $\sim$50$\%$ of clusters lying above the 10
Gyr isochrone. NGC~5846 and NGC~4365 show a steep drop in numbers between 1 and
5~Gyr and reach the 50$\%$ level already at an age of about 5~Gyr or
younger. This is consistent with the presence of a significant
fraction of younger objects.

\section{Conclusion}
The results presented here show that combined optical and  near-infrared photometry 
is a very promising method to detect intermediate-age stellar
populations. Given that already 2 out of 6 galaxies show a significant 
intermediate-age population this might revive the discussion on when
ellipticals formed the bulk of their stars.

%%%%%%%%%%%%%%%%%%%%%%%%%%%%%%%%%%%%%%%%%%%%%%%%%%%%%%%%%%%%

%%%%%%%%%%%%%%%%%%%%%%%%%%%%%%%%%%%%%%%%%%%%%%%%%%%%%%%%%%%%%%%%%%
\end{document}